\documentclass[aip,rsi,reprint,graphicx]{revtex4-1} 

\usepackage[pdftex]{graphicx}
\usepackage{graphicx}
\usepackage{amsmath}
\usepackage{color}
\usepackage{lineno}
\usepackage{epstopdf}
\usepackage{textcomp}
\usepackage{units}
\usepackage{verbatim}

\def\87Rb{$^{87}$Rb}

\begin{document}
\title{A simple laser locking system based on a field-programmable gate array}

\author{N. B. J\o rgensen}
\affiliation{Department of Physics and Astronomy, Aarhus University, Ny Munkegade 120, 8000 Aarhus C, Denmark}
\author{D. Birkmose}
\affiliation{Department of Physics and Astronomy, Aarhus University, Ny Munkegade 120, 8000 Aarhus C, Denmark}
\author{K. Trelborg}
\affiliation{Department of Physics and Astronomy, Aarhus University, Ny Munkegade 120, 8000 Aarhus C, Denmark}
\author{L. Wacker}
\affiliation{Department of Physics and Astronomy, Aarhus University, Ny Munkegade 120, 8000 Aarhus C, Denmark}
\author{N. Winter}
\affiliation{Department of Physics and Astronomy, Aarhus University, Ny Munkegade 120, 8000 Aarhus C, Denmark}
\author{A. J. Hilliard}
\affiliation{Department of Physics and Astronomy, Aarhus University, Ny Munkegade 120, 8000 Aarhus C, Denmark}
\author{M. G. Bason}
\affiliation{Department of Physics and Astronomy, Aarhus University, Ny Munkegade 120, 8000 Aarhus C, Denmark}
\affiliation{School of Physics \& Astronomy, University of Nottingham, Nottingham NG7 2RD, United Kingdom}
\author{J.~J.~Arlt}
\affiliation{Department of Physics and Astronomy, Aarhus University, Ny Munkegade 120, 8000 Aarhus C, Denmark}

\date{\today}

\begin{abstract}
Frequency stabilization of laser light is crucial in both scientific and industrial applications. Technological developments now allow analog laser stabilization systems to be replaced with digital electronics such as field-programmable gate arrays, which have recently been utilized to develop such locking systems. We have developed a frequency stabilization system based on a field-programmable gate array, with emphasis on hardware simplicity, which offers a user-friendly alternative to commercial and previous home-built solutions. Frequency modulation, lock-in detection and a proportional-integral-derivative controller are programmed on the field-programmable gate array and only minimal additional components are required to frequency stabilize a laser. The locking system is administered from a host-computer which provides comprehensive, long-distance control through a versatile interface. Various measurements were performed to characterize the system. The linewidth of the locked laser was measured to be $\unit[0.7\pm0.1]{MHz}$ with a settling time of  $\unit[10]{ms}$. The system can thus fully match laser systems currently in use for atom trapping and cooling applications.
\end{abstract}

\maketitle

\section{Introduction}
\label{sec:introduction}

Lasers are an essential technology in industry and scientific research. The frequency of the emitted light is usually determined by the laser gain profile and the length of the resonator cavity. Due to external perturbations such as temperature drifts or acoustic noise, the laser frequency drifts. For many applications, these drifts are not critical, but scientific and some advanced industrial applications often require precise frequency stabilization. Feedback circuits that stabilize the laser frequency have therefore been in use for decades. These circuits typically compare the laser frequency to a stable reference, e.g. by performing spectroscopy, and use a proportional-integral-derivative (PID) controller to provide a correction signal to control elements in the laser.

The field of atomic and molecular physics sets particularly strict demands on the laser frequency stability. To cool and trap atoms, lasers are typically frequency stabilized using an atomic transition or a high finesse cavity. Several different strategies have been developed to obtain a locking signal from a spectroscopic measurement\cite{wieman1976,shirley1982,overstreet2004,jundt2003,demtroder2003}. Often, the technique of saturated absorption spectroscopy combined with lock-in detection is used to produce a locking signal which is insensitive to non-frequency fluctuations of the laser~\cite{bjorklund1983,MacAdam1992,demtroder2003}.

Laser cooling and trapping experiments have become widespread in modern research laboratories and commercial products for the implementation of the required laser systems are available. The necessary components for such a system can, however, also be constructed in-house. Such an approach can be used to build low-cost setups to generate cold and ultracold atomic samples in undergraduate level laboratories~\cite{MacAdam1992,Wieman1995, arnold1998, Mellish2002, Singer2002, Whitaker2006}. A significant hurdle in such a setup is the construction of a laser frequency stabilization system which typically requires extensive analog electronics.

In this article, we present a laser locking system for frequency stabilization based on a field-programmable gate array (FPGA), which has several advantages compared to both analog electronics and commercial systems. It offers rapid reconfigurability, a user-friendly interface, it is low-cost and easy to contruct.

Several FPGA-based devices have been developed in the field of atomic and molecular physics~\cite{Restelli2005,Nikolic2013,Schwettmann2011,xu2012,Yang2012,leibrandt2015}, including various laser locking systems~\cite{Schwettmann2011,xu2012,Yang2012,leibrandt2015}. The first demonstration of a FPGA-based laser lock realized a side of fringe lock to a cavity~\cite{Schwettmann2011}. Later, a dedicated stand-alone system including parallel slow and fast feedback was contructed which achieved a linewidth in the kHz regime~\cite{Yang2012}. The most recent dedicated FPGA-based lock system was employed to stabilize a fibre laser to the precision required in an atomic clock~\cite{leibrandt2015}. Alternatively, a FPGA-based loop to optically phase-lock two lasers have been developed~\cite{xu2012}.

As an alternative to previous FPGA-based locks, we present a simple and versatile solution which is easy to implement. The system is fully applicable for challenging scientific applications such as cooling and trapping neutral atoms, but requires minimal additional electronics. It is based on a single off-the-shelf board which includes a FPGA, a microprocessor, analog-to-digital and digital-to-analog converters. Furthermore, this is the first FPGA-based lock programmed in LabVIEW, which significantly simplifies the task of understanding and reprogramming the software. This additionally allows for a versatile graphical user interface and long distance control.

The paper is structured as follows. In Sec.~\ref{sec:peaklock}, the procedure to obtain a locking signal is reviewed, followed by a description of the FPGA  implementation in Sec.~\ref{sec:implementation}. Optimization and characterization of the locking system is presented in Sec.~\ref{sec:performance} and conclusions are drawn in  Sec.~\ref{sec:conclusion}.

\section{Peak lock for atomic physics}
\label{sec:peaklock}

The laser locking system was developed for laser cooling and trapping of neutral atoms, where linewidths well below the typical atomic transition linewidths of approximately $\unit[5]{MHz}$ are required. This is achieved by stabilizing the laser frequency to a spectroscopy signal obtained in an atomic vapor cell. Since the Doppler-broadened signal provides insufficient frequency precision, Doppler-free saturation absorption spectroscopy is used~\cite{demtroder2003}.

\begin{figure}[t]
	\centering
	\includegraphics[width=8.5cm]{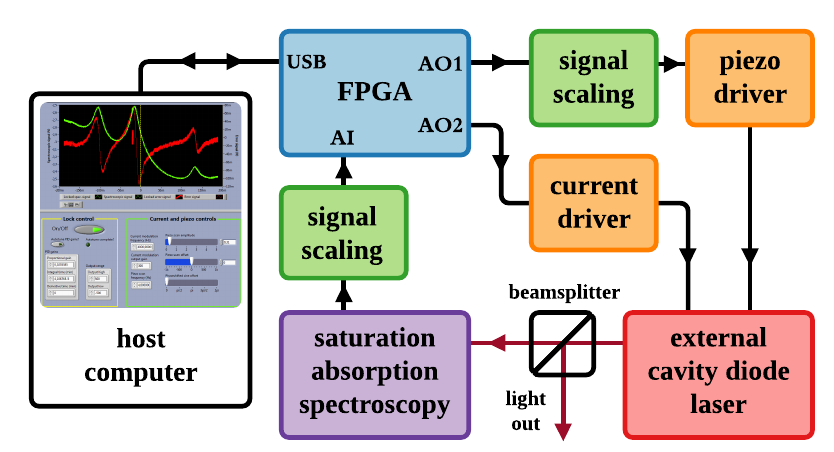}
	\caption{Overview of the feedback system with the main electronic signals (black lines) and light signals (red lines) displayed. The current and piezo drivers that supply the ECDL recieve their inputs from the FPGA. Laser light is sent to a saturated absorption spectroscopy setup, which produces the spectroscopy signal fed back to the FPGA.}
	\label{fig:setup}
\end{figure}

The design schematic of the feedback system is shown in Fig.~\ref{fig:setup}. The FPGA controls an External Cavity Diode Laser (ECDL)\cite{ricci1995} by providing input to its current driver and to the piezo driver that controls the position of the frequency selective grating inside the laser. The signal obtained from saturated absorption spectroscopy, is analyzed within the FPGA which then performs feedback accordingly.

To perform feedback, an error signal is required, which is proportional to the difference between a desired setpoint and the actual laser frequency. A PID controller attempts to minimize this error signal according to a proportional, an integral and a derivative term which respectively account for present, past and possible future values of the error signal. Each of the three terms contain a tunable coefficient, referred to a P, I and D parameters, which allow the strength of the terms to be adjusted.

The derivative of a spectroscopic signal provides a steep, linear zero crossing at an atomic resonance position and is thus well suited as an error signal. To obtain this derivative, heterodyne spectroscopy is employed~\cite{drever1983}. In this technique, the laser frequency is modulated with a frequency $\omega_\text{m}$, which produces two weak sidebands at $\pm \omega_\text{m}$. After passing through the vapor cell, the laser intensity $I$ is measured on a photodiode. In the limit of small modulation amplitude $M$, the detected light intensity is given by
\begin{align*}
I(t) = I_0 \left[1 + M \delta \cos (\omega_\text{m} t) + M \phi \sin (\omega_\text{m} t) \right],
\end{align*}
where $I_0$ is the intensity of the laser without modulation, $\delta$ is proportional to the difference in loss experienced by the two sidebands and $\phi$ is proportional to the difference between the phase shift experienced by the laser frequency and the average of the phase shifts experienced by the sidebands~\cite{bjorklund1983}. If $\omega_\text{m}$ is small compared to the width of the spectral feature, $\delta$ corresponds to the derivative of the spectroscopy signal and $\phi$ to the second derivative of the dispersion.

The demodulation of the signal is implemented through multiplication. An additional sinusoidal signal with frequency $\omega_\text{m}$ and phase $\theta$ is introduced and multiplied with the spectroscopy signal
\begin{align*}
I(t) \cos(\omega_\text{m} t + \theta) & = \frac{ I_0 M}{2} \bigg[ \delta \cos (\theta) - \phi \sin (\theta)  \\ + \frac{2}{M} \cos (\omega_\text{m} t + \theta)
& + \delta \cos (2 \omega_\text{m} t + \theta)  + \phi \sin (2 \omega_\text{m} t + \theta)  \bigg].
\end{align*}
The resulting signal contains three terms which depend on $\omega_\text{m}$; in practice these are eliminated by a low-pass filter. Finally, by adjusting $\theta$ to zero, $\phi$ is eliminated and the derivative $\delta$ is obtained and available for feedback. 

\section{FPGA implementation} 
\label{sec:implementation}

Our laser frequency stabilization system is based on a National Intruments myRIO-1900 device, which is a hardware platform containing a programmable FPGA, a microprocessor, analog-to-digital and digital-to-analog converters\footnote{The datasheet of the myRIO device is available online at http://www.ni.com/pdf/manuals/376047a.pdf}. The device can easily be accessed and programmed using LabVIEW software. No custom FPGA functions have been used in our implementation, making the software highly transferable to other LabVIEW based FPGA systems. The source code of the software used in this work is available online along with future versions~\footnote{Updated source code is available online at http://phys.au.dk/ forskning/forskningsomraader/uqgg0/downloads/}.

\begin{figure}[tb]
	\centering
	\includegraphics[width=8.5cm]{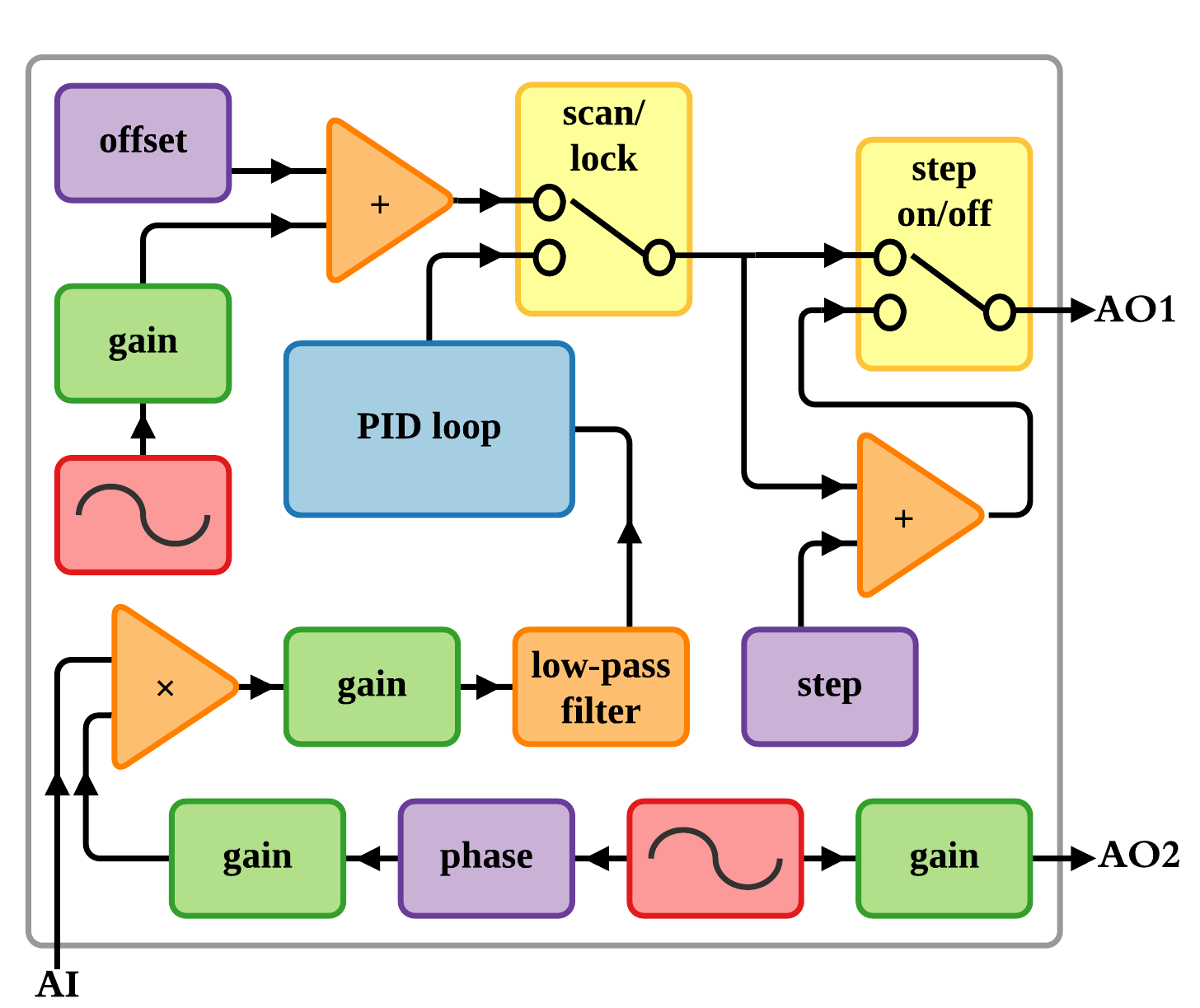}
	\caption{Dataflow within the FPGA. Before the signal enters the FPGA, it is digitized to a 12 bit signal, which is further upconverted to 16 bit within the FPGA. Finally, all signals are converted to 12 bit before digital-to-analog conversion. The signal paths are further explained in the text.}
	\label{fig:FPGA}
\end{figure}

\begin{figure*}[tb]
	\centering
	\includegraphics[width=17cm]{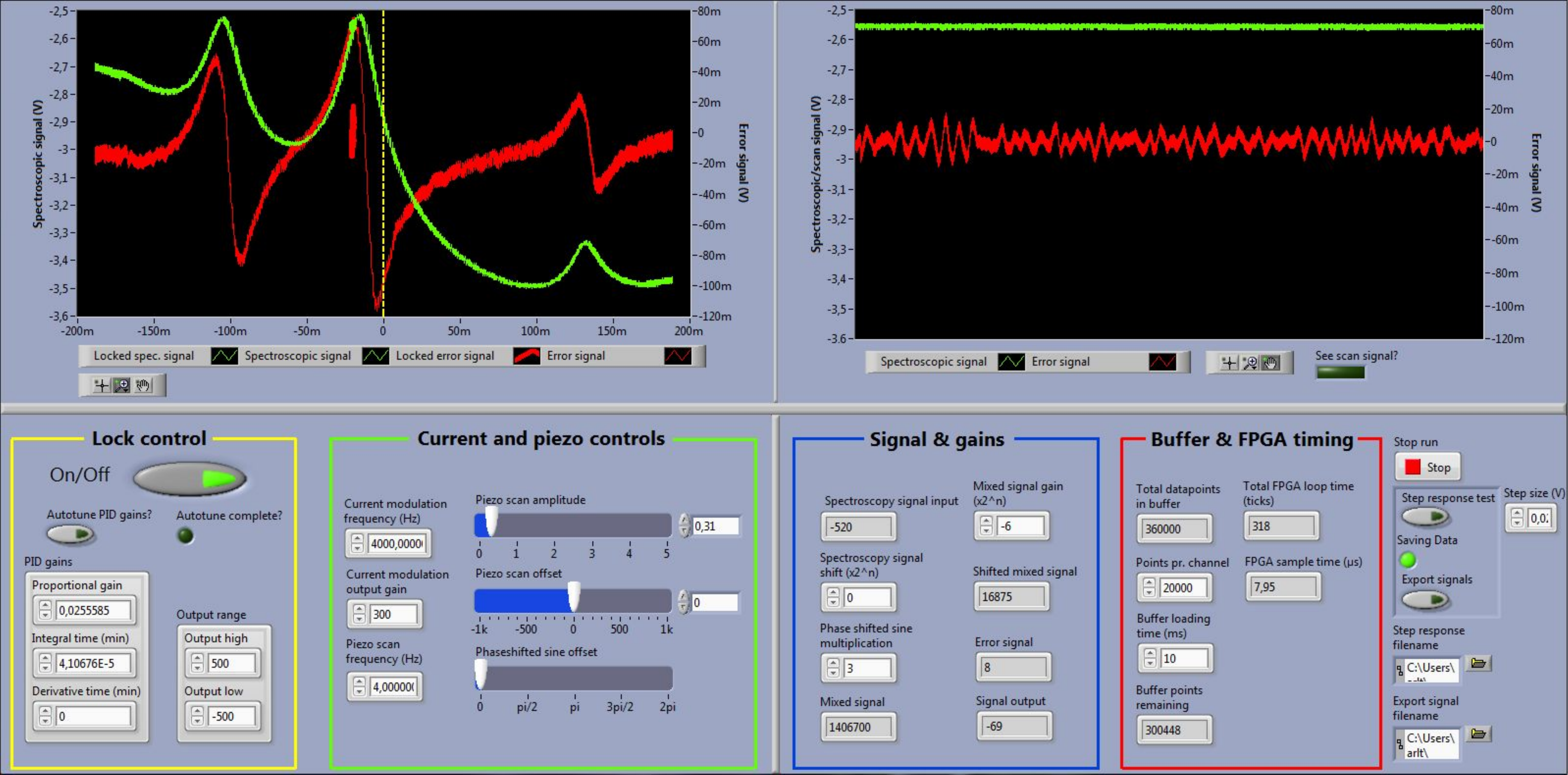}
	\caption{User interface in locked mode. The interface is divided into four panels. The upper left panel shows the spectroscopic signal (green) and the error signal (red) versus the piezo scan value. In locked mode, the latest scan image is frozen and the PID loop is active. In this case, the spectroscopic signal and error signal are displayed on top of the frozen image. The central peak displayed is the crossover peak between the $F=2 \rightarrow F'=2$ and $F=2 \rightarrow F'=3$ transitions of the $^{87}$Rb-D2 line, which the laser is locked to. The upper right panel displays the spectroscopic and error signal for a timespan of $\unit[160]{ms}$, which is convenient to track the lock situation. The locking controls are located in the lower left panel where current modulation, piezo scan and PID parameters can be chosen and the status can be switched between lock and scan mode. The lower right panel displays the internal FPGA signals and additional gains can be adjusted. In the far right corner, there is a panel to export data.}
	\label{fig:interface}
\end{figure*}

The FPGA delivers the input to the current and piezo drivers, receives a signal from the spectroscopy setup and performs the feedback. The frequency modulation is realized by modulating the diode current, and the feedback is performed by adjusting the piezo voltage. The data flow within the FPGA is outlined in Fig.~\ref{fig:FPGA} and discussed in the following paragraphs.

The laser used in the setup is a home-built ECDL\cite{ricci1995}. It delivers light at $\unit[780]{nm}$ and the spectroscopy and stabilization is performed using a rubidium vapour cell containing the two isotopes $^{87}$Rb and $^{85}$Rb in their natural abundances. In the following, the laser is locked to the crossover peak between the $F=2 \rightarrow F'=2$ and $F=2 \rightarrow F'=3$ transitions of the $^{87}$Rb-D2 line\footnote{A crossover peak is located between two regular transitions and is a consequence of a specific non-zero velocity class of atoms being simultaneously resonant with the counterpropagating pump and probe beams}, but any other spectroscopic feature could equally well be chosen.

The locking system is controlled from a host computer using an interface which displays all relevant data as shown in Fig.~\ref{fig:interface}. The modulation of the laser current to produce sidebands is generated by a digital sine wave generator and sent to the analog output AO2. The spectroscopy signal is scaled to match the voltage range of the analog-to-digital converter (see Fig.~\ref{fig:setup}), then it is digitized, and sent to the FPGA via the input channel AI. It is then multiplied with the phase-shifted modulation signal and by applying a filter, the error signal has been obtained. The low-pass filter used is a 4th order Butterworth filter implemented with a cut-off frequency of $\unit[500]{Hz}$, removing the modulation signal of several kHz. Subsequently, the error signal enters the PID controller, which then produces a feedback signal accordingly.

It is often necessary to scan the laser over a range of frequencies to locate and resolve the spectroscopic features for locking. Thus, two different modes of operation are required: a locked mode and a scan mode. In scan mode, a broad range of frequencies is scanned and in lock mode, the feedback stabilizes the laser frequency to a given feature.

When the system is in scan mode, a piezo voltage scan signal is generated by a second digital sine wave generator and sent to the output channel AO1 which supplies the input signal to the piezo driver. In locked mode, the piezo signal is supplied by the PID controller. A key property of the feedback system is its response to a sudden disturbance. A switch has been implemented which can add an offset to the piezo signal, allowing a straightforward test of the step response of the system.

In order to display the obtained data and control the system, the spectroscopy signal, the error signal and the piezo signal is transferred from the FPGA to the host computer. Since the myRIO only has two built-in DC-coupled analog outputs, the AC coupled stereo audio output channel is utilized to allow independent external monitoring of the error signal.

\section{Lock Performance}
\label{sec:performance}

In this section, the optimization and characterization of the locking system is described. This includes a measurement of the system response as a function of the modulation frequency and the implementation of a step response technique, which allows for an optimization of the settling time. Furthermore, the locking performance is determined by measuring the power spectral density of the frequency fluctuations and the linewidth of the laser through beat measurements.

The primary goal of the laser lock is to obtain fast recoveries from disturbances. This is dependent on the steepness of the error signal slope which in turn depends on the modulation frequency. Since the modulation frequency sets the upper limit for the bandwidth of the locking circuit, a low modulation frequency is undesirable. Additionally, the absorption difference experienced by the sidebands is lower for small modulation frequencies. At high frequencies, technical limitations such as the bandwidth of the system originating from the speed of the FPGA are relevant.

\begin{figure}[tb]
	\centering
	\includegraphics[width=8.5cm]{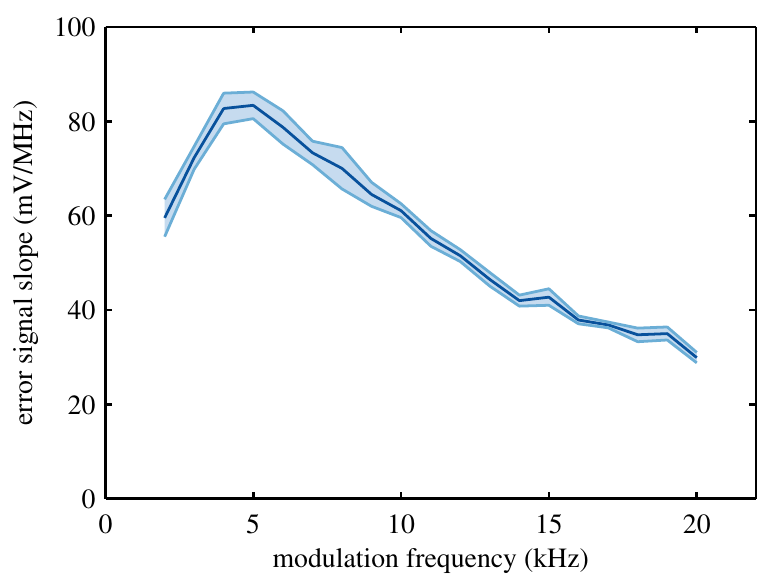}
	\label{fig:slopefreq}
	\caption{Slope of the error signal versus modulation frequency. The shaded area is the standard error of the mean for five measurements.}
	\label{fig:slopeVSfreqMod}
\end{figure}

To optimize the slope of the locking signal, it was measured for a range of modulation frequencies with constant modulation amplitude as shown in Fig.~\ref{fig:slopeVSfreqMod}. Each measurement was repeated five times and the slope was found using a linear fit. The highest locking signal derivative is obtained at $4$--$\unit[5]{kHz}$.

The error signal slope was also measured as a function of the modulation signal amplitude, which displayed a linear dependence for small and moderate amplitudes. A large slope is preferable for the system to react to disturbances, but the linewidth grows at large modulation amplitudes. Hence, depending on the purpose of the laser, the modulation amplitude should be adjusted accordingly.

In the first experiments with the locked laser, the PID parameters were optimized using the step response technique. The Nichols-Ziegler method~\cite{ziegler1942} served as a starting point. Subsequently, the parameters were adjusted to minimise the step response time. A disturbance to the locked laser was introduced through the step switch programmed into the FPGA (see Fig.~\ref{fig:FPGA}). The step response time was defined as the time it took from the beginning of the disturbance until the spectroscopy signal returned to its locked value $\pm 1 \sigma$ of the noise level. 

\begin{figure}[tb]
	\centering
	\includegraphics[width=8.5cm]{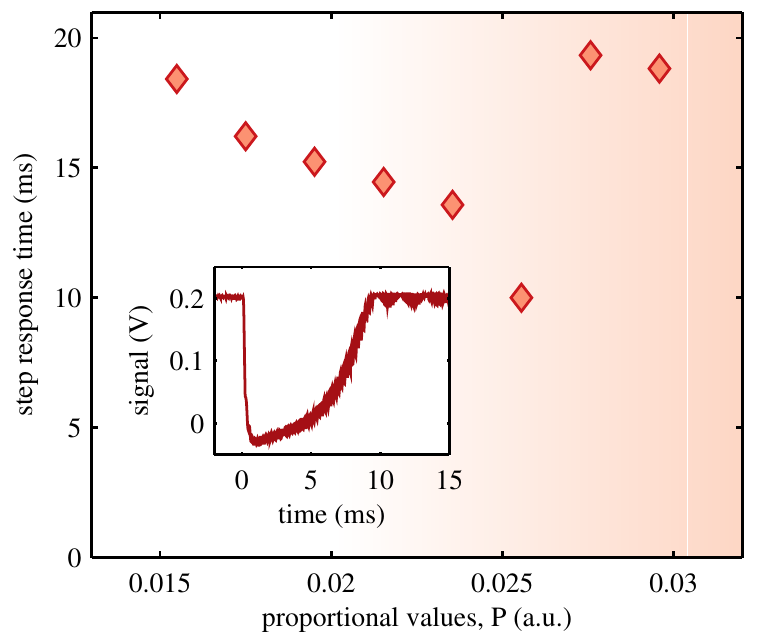}
	\caption{Settling time for different values of the P parameter. The shaded area indicate the appearence of oscillations. The inset shows a spectroscopy signal for the minimum step response time of $\unit[10]{ms}$.}
	\label{fig:PIDopt}
\end{figure}

Part of this optimization is shown in Fig.~\ref{fig:PIDopt}, where the step response time was recorded for different values of the P parameter for constant I and D parameters. Each data point shows the average of five measurements. It is evident that the response time decreases with increasing values of P, until substantial oscillations set in. The response for the optimum value is displayed in the inset figure, where the spectroscopy signal settles in $\unit[10]{ms}$. 

To determine the performance of the laser locking system, the power spectral density of frequency fluctuations was measured in unlocked and locked mode. The measurement was performed by recording a $\unit[5]{s}$ trace of the error signal digitally within the FPGA at a sample rate of $\unit[125]{kHz}$. In both cases the frequency of the laser was tuned to the crossover peak of $^{87}$Rb.

\begin{figure}[tb]
	\centering
	\includegraphics[width=8.5cm]{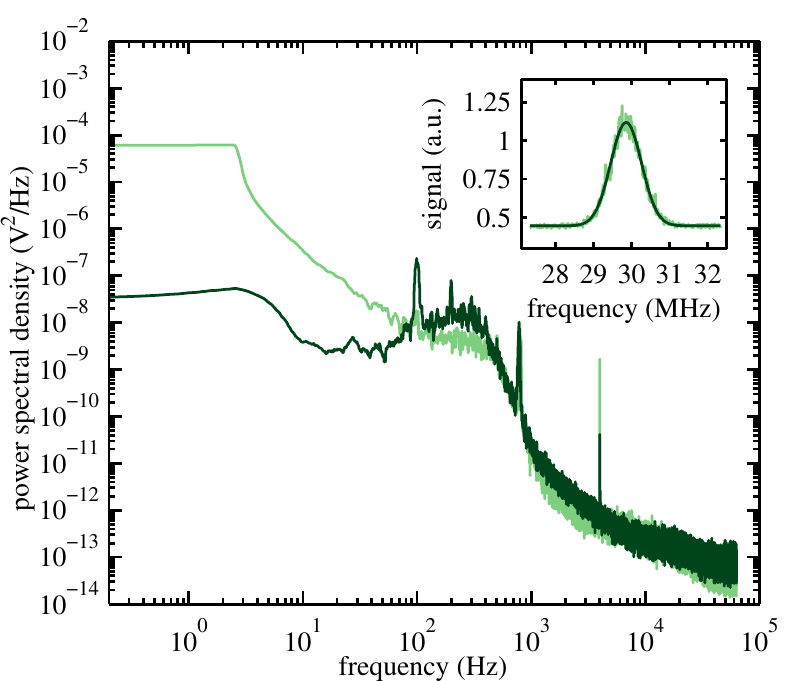}
	\caption{Power spectral density of frequency fluctuations of the laser when unlocked (light green line) and when locked (dark green line). The inset shows one of the beat measurements (light green), which is used to determine the linewidth of the laser by applying a Gaussian fit (dark green line).}
	\label{fig:noisespectrum}
\end{figure}

The results are displayed in Fig.~\ref{fig:noisespectrum} where the Fourier transformation of the signal has been taken. In frequency space, each data point is averaged with its 50 nearest neighbours to give a good estimate of the true mean value in each segment. A clear damping of the noise is observed at lower frequencies when locked, demonstrating the ability to correct for frequency drifts. A noise increase is seen at $\sim \unit[300]{Hz}$, corresponding to the bandwidth of the feedback system. The narrow peak at $\unit[100]{Hz}$ is electronic noise and the current modulation of $\unit[4]{kHz}$ is also visible in the spectrum. It is likely that the $\unit[800]{Hz}$ peak orignates from the myRIO device which contains an accelerometer that operates at that sample rate.

The observed frequency noise can be compared with the step response signal shown in the inset of Fig.~\ref{fig:PIDopt}. The bandwidth of some $\unit[100]{Hz}$ matches the step response time of $\unit[10]{ms}$ and the high frequency noise of the spectroscopy signal originates from the $\unit[4]{kHz}$ modulation. 

The oscillations observed after $\unit[10]{ms}$ are damped further on longer time scales and correspond to the bandwidth of the feedback. Depending on the choice of PID parameters, a faint oscillation remains due to a slight overcompensation of the feedback. Usually, the oscillations are not visible in the spectroscopic signal, but only in the error signal as can be observed in the right panel of Fig.~\ref{fig:interface}.

The long-term stability of the system was tested by leaving it untouched, and in this case the laser remained locked for up to one day. The main limitation to this stability is the mechanical construction of the ECDL.

Finally, the linewidth of the laser was determined. A series of beat measurements was performed with two other lasers of similar design and wavelength. Those lasers are currently in use for the production of $^{87}$Rb Bose-Einstein condensates~\cite{pedersen2014} and locked using home-built analog PID controllers acting on a locking signal derived from a $\unit[10]{MHz}$ modulation of the laser current. For these measurements, all lasers were locked to the crossover peak of $^{87}$Rb and acousto-optical modulators were used to shift each pair of lasers to be separated in frequency by $\approx \unit[30]{MHz}$. For each measurement, the beams of two lasers were coupled into a single-mode polarization maintaining fiber to ensure good spatial mode overlap. The fiber output was then directed onto a photodiode and the beat signal was obtained using a spectrum analyzer which recorded spectra at sweeptimes of $\unit[500]{ms}$.

An example of such a beat measurement is shown in the inset of Fig.~\ref{fig:noisespectrum}. The beat spectrum is composed of a Lorentzian profile due to high frequency noise of the laser and a Gaussian profile due to the drift of the laser within an individual measurement. The signal to noise ratio of our specific setup only allowed the Gaussian component to be resolved. This is fully sufficient, since the Gaussian component determines the effective linewidth in these types of laser systems.

The linewidth of the FPGA-based laser system was found to be $\unit[0.7\pm0.1]{MHz}$ using low modulation amplitudes. The other two lasers tested were found to have linewidths of $\unit[0.4\pm0.1]{MHz}$ and $\unit[0.6\pm0.1]{MHz}$. No signi\-ficant alterations of the linewidths were observed for the range of sweeptimes $100$--$\unit[500]{ms}$. Thus, our new laser locking system performs similarly to laser systems currently in use for experiments with ultracold atoms.

\section{Conclusion}
\label{sec:conclusion}

We have developed a simple FPGA-based laser locking system for frequency stabilization. It was optimized and characterized in a series of measurements, which demonstrates the flexibility of the system. The measured laser linewidth was $\unit[0.7\pm0.1]{MHz}$, similar to linewidths of lasers which are currently in use for laser cooling and trapping experiments.

Compared to initial results from FPGA-based locks, this is an improvement by more than a factor of two~\cite{Schwettmann2011}. On the other hand, our laser-lock does not achieve linewidths comparable to dedicated locks for high precision experiments~\cite{Yang2012,leibrandt2015}.

In contrast, our objective was to develop a simple laser locking solution based on low-cost off-the-shelf hardware, which requires minimal effort to implement in any setting, yet still acheives a performance sufficient for complex tasks such as cooling and trapping of neutral atoms. It was programmed using LabVIEW which is a standard experimental control language and thus the system is easy to understand, construct and configure and is applicable in both state of the art laboratories, teaching environments and remote sensing applications.

\section{Acknowledgments}

We thank the Danish Council for Independent Research, the Lundbeck Foundation, the Templeton Foundation, and a Marie Curie IEF in FP7 for support.

\bibliography{FPGApaper}
\end{document}